\documentclass[aps,prl,twocolumn,showpacs, nofootinbib]{revtex4-1}  
\usepackage{amsmath, amssymb}        
\usepackage{graphicx, multirow, color}

\newcommand{\tev}{~\text{TeV}}
\newcommand{\gev}{~\text{GeV}}
\newcommand{\met}{\ensuremath{{\not\mathrel{E}}_T}}
\newcommand{\mptvec}{\ensuremath{{\not\mathrel{\vec{p}}}_T}}

\begin{document}
\title{Chasing New Physics in Stacks of Soft Tracks}
\author{Amit Chakraborty, Sabyasachi Chakraborty and Tuhin S.~Roy}
\affiliation{Department of Theoretical Physics, Tata Institute of 
Fundamental Research, Mumbai 400005, India}

\date{\today}
\begin{abstract}
In this letter we introduce a new variable $\xi$, namely the number of tracks associated with the primary vertex, which are not parts of reconstructed objects such as  jets/isolated leptons etc. We demonstrate its usefulness in the context of new physics searches in the channel  monojet$+$missing transverse  momentum ($\met$).  In models such as in compressed supersymmetry, events are often characterized by a rather large number of soft partons from the cascade decays, none of which result in reconstructed objects. We find that $\xi$, binned in  $p_T$, can discriminate these new physics events from events due to $Z+\text{jets}$, that is, the main background in the channel monojet$+ \met$. The information contained in soft tracks is largely uncorrelated with traditional variables such as the effective mass,  $\met$, $p_T$ of the jet, etc. and, therefore, can be combined with these to increase the discovery potential by more than  $200\%$ (depending on the spectra, of course).  In fact, we find that simple cuts on $\xi(p_T)$ along with cuts on $\met$, and the effective mass outperforms sophisticated  optimized MultiVariate Analyses using all conventional variables.  One can model the background distribution of $\xi(p_T)$  in an entirely data-driven way,  and make these robust against pile-up by identifying the primary vertex.  
   
\end{abstract}

\maketitle


The search for new physics (NP) in the monojet channel provides an intriguing phenomenological challenge. The observed event topology is simple:  a solitary jet with a large transverse momentum (namely, $p_T^J$) recoils against no other reconstructed objects. At the level of reconstructed objects,  the four-vector of the observed jet is the only observable. At the level of detector objects, there exists more information.  Energy deposited in various parts of the calorimeters, and/or tracks seen at the tracker and at the muon spectrometer result in the measurement of the  missing transverse momentum ($\mptvec$) and missing energy ($\met = \left| \mptvec \right|$). The background is simplistic:  enforcement of  the observation of \textit{one and only one} jet in the events implies that one mainly needs to worry about QCD and $W/Z+\text{jets}$. Ensuring that the jet momentum is not aligned with the $\mptvec$ (\textit{i.e.}, $\met$ is not dominantly due to energy mis-measurement), QCD can be easily taken care of.  Unfortunately, suppressing $Z(\nu\nu)+\text{jets}$ is hard (especially for the $p_T^J \sim \met \sim \mathcal{O}(100\gev)$).  The main difficulty arises from the fact that the information available is minimalistic in nature.  In conventional analyses a myriad of other variables in the form of various scalar and vector sums of visible particle momenta, are often being considered. However, these turn out to be highly correlated, and improving $S/B$, or more importantly,  $S/\sqrt{B}$  in this channel remains both difficult and challenging.
 
A plethora of  well-motivated models exist, where events due to NP show up in the monojet channel with moderate $\met$ and $p_T^J$, the most consequential of which are models of electroweak supersymmetry with a compressed spectrum~\cite{Martin:2007gf,Fan:2011yu,Murayama:2012jh}. As data from the Large Hadron Collider (LHC) keeps pouring in and, as a result, exclusion contours keep striding deep into the parameter space of the Minimal Supersymmetric Standard Model, models with  compressed spectra might very well turn out to be the last vestige of naturalness\footnote{Exceptions to the statement include models of neutral naturalness~\cite{Chacko:2005pe, Burdman:2006tz, Craig:2015pha}, various neat constructs in case of supersymmetry~\cite{Roy:2005hg, Csaki:2005fc, Falkowski:2006qq,  Perez:2008ng, Nelson:2015cea,  Cohen:2015ala} etc.}. Note that the signal topologies in compressed scenarios are the same as in traditional hierarchical spectra, where colored superpartners produced at the top of the  decay chain cascade down to the collider stable Lightest Supersymmetric Particle (LSP), giving rise to a bunch of partons, leptons, etc. Because of the small mass differences between the colored superpartners and the LSP in compressed scenarios, all visible particles end up being rather soft -- too soft to give rise to reconstructed objects, or even a large $\met$.  Indeed, in the limit gluino and the LSP are well separated, the bound on degenerate squark-gluino masses has already reached $1.8\tev$~\cite{Aad:2014wea}, whereas  in compressed cases we have failed to exclude beyond even $600\gev$~\cite{Aad:2015zva}. 

Note that the inability to probe compressed spectra, in some way, reveals an essential limitation in collider physics: every reconstructed object (whether that is an isolated electron, muon, photon, or even a jet) has a minimum $p_T$ associated with it, which is considerably higher than the $p_T$ required for a particle to be recorded as a detector object. The tracker, for example, may record a $\pi^{+}$ with $p_T \sim \text{ few}\gev$ (resulted in the shower and subsequent hadronization of a parton emanating from the short distance hard process), but for the $\pi^{+}$ to give rise to a reconstructed object (in this case, a jet), it requires other detectors objects surrounding it to satisfy the $p_T$ criteria collectively. Compressed spectra provide an excellent case study where this point is highlighted, and, therefore, should be studied in detail -- irrespective of the existence of any theoretical motivation. Finding non-standard search strategies to discover these spectra, unavoidably improves the reach of LHC. 

It is clear that in order to probe an event originating due to a compressed spectra, one needs to use the detector information directly: analogous to the jet-substructure studies which deal with detector level information but only within a jet. To be fair, the $\mptvec$, the scalar transverse energy sum ($H_T$), or the effective mass $M_\text{eff} = H_T + \left| \mptvec \right|$ do so already by definition. 
\begin{equation}
\mptvec = - \sum_{i} \vec{p}_{T}^{i}   \, , \quad
H_T =  \sum_{i} \left| \vec{p}_{T}^i \right|   \, , 
\end{equation}
where $i$ refers to the detector level objects (such as particle flows, or tracks and calorimeter cells). Broadly speaking, however, all these observables are $p_T$ or energy weighted and are, therefore, highly correlated. Even an optimized MultiVariate Analyses (MVA) fails to enhance the discovery potential significantly using only these. A big enhancement in signal significance in this channel would necessitate finding observables that are somewhat uncorrelated to the existing set. 

The central idea of this letter stems from the observation that none of the existing variables give a measure of the particle multiplicity in the event. A simple counting of tracks does carry that information and, therefore, is expected to be mostly uncorrelated to any of the other variables mentioned in this work. However, one needs to be careful in order to deal with tracks. The number of charged hadrons resulted in a shower is an infrared unsafe quantity; cuts on the number of tracks might give rise to unexpected scales in the event shapes~\cite{Martin:2016jdw}; tracks inside a reconstructed object mostly carry information regarding the object itself; and there are contaminations due to underlying events and pile-up. 

In this letter we advocate for the use of the number of tracks  that are $(i)$ associated with the primary vertex, and $(ii)$ are not part of any reconstructed object. By this we mean the angular separation between the jet and the tracks is greater than the size of the jet. Further we bin them according to their $p_T$, namely $ p_T  > 5\gev$,  $1\gev < p_T  \leq 5\gev$, and finally $0.5\gev < p_T  \leq 1\gev$. We denote this variable by $\xi(a)$, where  $a$ is one of $\{ 5, 1, 0.5 \}$, designating the left boundary of the bins in $\gev$.  Note that a cut on $p_T$,  somewhat ameliorates the problem of arbitrary soft splitting in a shower. Identification of the primary vertex in the event and only counting the tracks associated with it, gives $\xi$ robustness against pileup. Finally, tracks \textit{outside} the reconstructed objects carry mostly knowledge of the full event\footnote{Other non-standard ideas that can be used in the context of compressed spectra include the use of soft isolated leptons~\cite{Giudice:2010wb}, tagging of soft but not ISR jets~\cite{Delgado:2016gqn}, or looking for disappearing tracks for long-leved particle scenarios~\cite{Aad:2013yna}.}.

Let us reiterate that the point of this paper is not to discuss the discovery potential of some well-motivated spectra, but rather to add a new \emph{tool} and to quantify its effectiveness.  In order to achieve this we take a sample compressed spectrum and calculate $S/B$ and $S/\sqrt{B}$ combining $\xi $ with the conventional variables. Since we are only interested in the performance of $\xi$,  all our results will actually be double ratios, where we only show enhancements w.r.t the performance of an optimized MVA that uses all the conventional variables.  The variables, therefore, are  grouped in two sets, a set with only the conventional variables (namely, $V_\text{c}$), and another set where the $\xi$ variables are included  (namely, $V_\text{all}$). To be specific:
\begin{equation} 
\begin{split}
V_\text{c} \ & \equiv \ \{ p_T^J, \met, H_{T}, M_\text{eff} \} \\
V_\text{all}\ & \equiv \ \{\xi(5), \xi(1), \xi(0.5),  p_T^J, \met, H_{T}, M_\text{eff} \}
\end{split}
\label{variables}
\end{equation}

\begin{figure}[htb!]
\center
\includegraphics[width=0.4\textwidth]{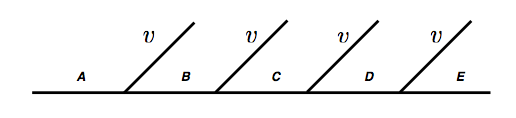}
\caption{The decay chain used in the analysis. In the figure, $v$ refer to visible remnants of a decay (such as partons, charged leptons etc.), none of which give rise to jets/isolated leptons, but contribute to $\xi$. }
\label{fig:topo}
\end{figure}

As far as the event topology is concerned, we pair produce a hypothetical particle $A$ associated with a hard parton (we use $AAj$ to refer to this process), which goes through a cascade of decays as shown in Fig.~\ref{fig:topo}, giving rise to a set of visible particles and a set of invisible particles (denoted in the figure collectively by $E$). Note that we use a long decay chain, since a larger number of soft decay products will  enhance $\xi$  for the $AAj$ events. 
In this work, we use $m_A-m_B = m_B-m_C= m_C-m_D = m_D-m_E= 25\gev$, with $m_A = 1.5\tev$. This results in the degree of compression $\Delta \equiv  \left( m_A - m_E\right)/m_A=1/15$, a typical value for the compressed spectra used in literature~\cite{Baer:2007uz,Martin:2008aw,LeCompte:2011fh,Bhattacherjee:2013wna,Mukhopadhyay:2014dsa,Baer:2014kya,Han:2014kaa,Harigaya:2014dwa,Ellis:2015vaa,Nagata:2015hha,Bramante:2015una, Han:2015lha,Dutta:2015exw}. In our analysis, to be specific, we use $A, B, C, D, E$ to be $\widetilde q$, $\widetilde g$, $\widetilde t$, $\widetilde\chi^+$ and $\widetilde\chi^0$ respectively in order to generate signal events. However, we note that $\xi$ is expected 
to be equally effective in models with a large number of soft particles, for example $R$-parity violating SUSY \cite{Barbier:2004ez}, models with universal extra dimensions consisting of degenerate $KK$-modes \cite{Georgi:2000ks,Cheng:2002iz,Murayama:2011hj} etc.

For the signal events we produce $AAj$ events at the matrix element level.  The Standard Model (SM) processes that can contribute to our monojet + $\met$ topology include $W/Z+\text{jets}$,  $t\bar{t}$, and QCD. In principle, single $t+\text{jets}$ and  $\text{di-}W/Z +\text{jets}$ can also contribute. As we mention earlier, simple event preselection criteria can suppress all background except $Z(\nu\bar{\nu})+\text{jets}$ and $W(l\nu)+\text{jets}$ (if the lepton is missed).  Since the events topologies for these electroweak events are identical, we expect $\xi$ to be equally effective in suppressing $Z/W+\text{jets}$.  In this letter, we are only interested in demonstrating the performance of $\xi$, and therefore, we only consider $Z$ produced with a single hard parton at the matrix element level (referred to here by $Zj$ events), which subsequently decays to neutrinos.

We generate both signal and background events using {\tt Madgraph\!\! 5}~\cite{Alwall:2014hca} with a cut on the associated parton momentum $\hat{p}_T > 50\gev$ for $pp$ collisions at the center of mass energy of $13\tev$. For hadronization and showering we use {\tt PYTHIA\!\! 8.2}~\cite{Sjostrand:2014zea} with parton distribution function~{\tt CTEQ-6}~\cite{Pumplin:2002vw}. We also utilize the default underlying event modeling as implemented in {\tt PYTHIA}.  In order to perform a semi-realistic detection simulation, we use {\tt Delphes\!\! 3.3}~\cite{Ovyn:2009tx,deFavereau:2013fsa} with the default CMS card. For simulating pileup, we generate low-$Q^2$ soft QCD events using {\tt PYTHIA}. Mixing of these pile-up events with the events due to hard interactions  are subsequently performed using {\tt Delphes}. We use default parametrization as implemented in the CMS card, in order to randomly distribute pile-up and hard scattering events in time and $z$ positions. The number of soft events merged with each hard scattering follows Poisson distribution with a mean (say, $\langle N_\text{PU}\rangle$ ).  In this work we consider $\langle N_\text{PU}\rangle$ to be $20$ and $40$. {\tt Delphes} also provides modules to subtract both neutral and charged pileup.  After identifying the primary vertex, we use {\tt Delphes} to remove all tracks for which $|z|$ is larger than the spatial vertex resolution of the tracker (we use the default value of $0.25$). Finally,  we record detector outputs in terms of tracks and particle flow momenta separately. Jets are constructed from the particle flows using anti-$k_T$ jet algorithm~\cite{Cacciari:2008gp} with  $R=0.5$ and $p^\text{min}_{T} =  50\gev$. We keep events with exactly one jet with $p_T^J>100\gev$ in the central part of the detector ($\left| \eta \right| < 2.0$), and with no other reconstructed objects for further processing. This is our preselection criteria, and  all results presented in  this paper are based on events that satisfy these requirements. 

\begin{figure}[!h]
\centering
\includegraphics[width=0.5\textwidth]{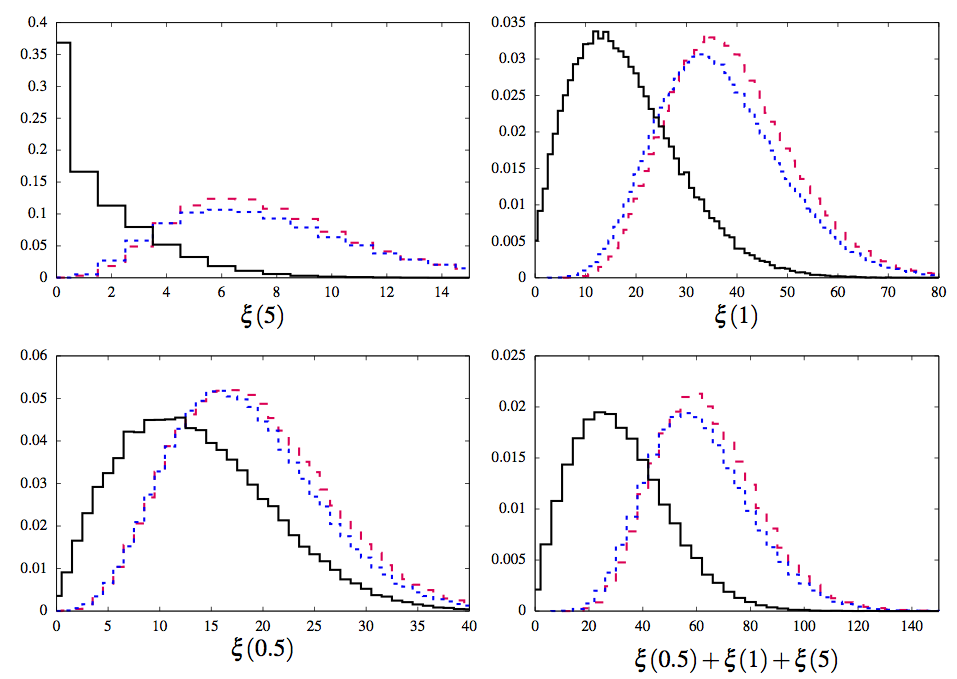}
\caption{\label{fig:xi} PDFs of $\xi(a)$ for various values of $a$ as described in the text for $\langle N_\text{PU}\rangle = 40$. The lower right frame shows the PDF when tracks in all the three bins are considered. The solid black lines in these plots show the PDFs for $Zj$ events, whereas,  red dashed and blue dotted figures show the same for $AAj$ and $BBj$ events respectively.  }
\end{figure}
In Fig.~\ref{fig:xi} we show the distribution of $\xi$ for both background and signal events for $\langle N_\text{PU}\rangle = 40$ subject to a $p_T$ cut as already discussed in the text. The right most figure in the lower half of Fig.~\ref{fig:xi} depicts the total number of charged tracks in an event which are not associated with the reconstructed jet.  As evident in these plots, $\xi$ clearly distinguishes between $AAj$ events (red dashed) and $Zj$ events (solid black).  For demonstration purposes  we also show $\xi$ distributions for events with $BBj$ (blue dotted). A clear correlation is apparent: the number of visible particles associated with the short distance processes increases from $Zj$ to $BBj$ to $AAj$, the same behavior is observed in any of the $\xi$ plots.   

\begin{figure}[!h]
\centering
\includegraphics[width=0.47\textwidth]{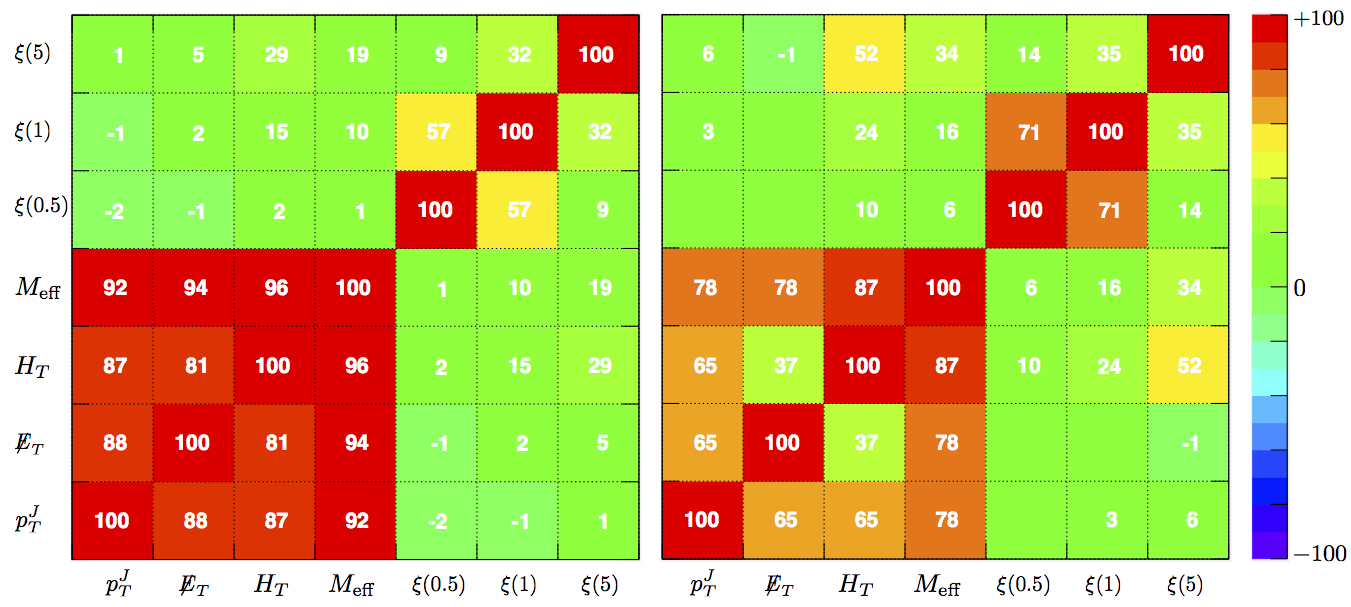}
\caption{\label{fig:corr} Linear correlation coefficients (in $\%$) for the $AAj$ events (left), and the $Zj$ events (right) for all the variables discussed in this paper.}
\end{figure}

Even though, the plots in Fig.~\ref{fig:xi} show that $\xi$ can be useful in separating signal from background,  these do not exactly reveal whether $\xi$ carry extra information over the conventional variables. For this we study the linear correlation coefficients $\rho\left( x,y \right) = \text{Cov}\left( x,y \right)/ \sigma_x\sigma_y$ for all pairs of variables. The coefficients for both signal and background are shown in the left and the right frames in Fig.~\ref{fig:corr} respectively. 
Note that a large linear correlation (anti-correlation) between two variables results in $\rho =  +1(-1)$ respectively. As can be seen from Fig.~\ref{fig:corr}, all conventional variables are highly correlated  in line with our claim. A striking feature of these plots are that these clearly show lack of linear correlations between $\xi$ and the conventional variables. In fact, taking the cue from Fig.~\ref{fig:corr}, we can define a minimal set of selective variables that should be able to outperform the conventional variables.  
\begin{equation} 
V_\text{sel}\  \equiv \ \{\xi(5), \xi(1), \xi(0.5), M_\text{eff} \}
\end{equation}

To quantify the full impact of $\xi$ as discriminating variables, we resort to MVAs using the Boosted Decision Tree (BDT) algorithm as implemented in the Toolkit for Multivariate Data Analysis~\cite{Hocker:2007ht} in the ROOT framework \cite{Antcheva:2009zz}. The parameters associated with the BDT analyses are chosen as follows:  the number of trees in the forest {\tt MaxDepth}= $400$, the maximum depth of the decision tree {\tt MaxDepth} $=5$,  and finally, the minimum percentage of training events required in a leaf node {\tt MinNodeSize}  $=2.5\%$. We keep all other necessary variables at there default values. Moreover, we consider the {\tt AdaBoost} method for boosting the decision trees in the forest with the boost parameter {\tt AdaBoostBeta} $=0.5$. 

\begin{figure}[!h]
\centering
\includegraphics[width=0.4\textwidth]{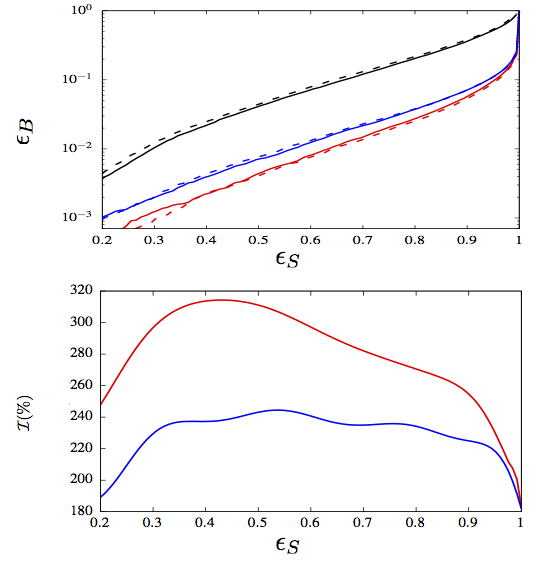}
\caption{\label{fig:roc1} Top: $\epsilon_B$ as a function of $\epsilon_S$ for $\langle N_\text{PU} \rangle = 40$ (solid) and $\langle N_\text{PU} \rangle = 20$ (dashed) using the three set of variables $V_C$ (black), $V_{\text{sel}}$ (blue) and $V_{\text{all}}$ (red). Bottom: $\mathcal{I}$ (in $\%$) as a function of $\epsilon_S$  for the two sets of variables $V_{\text{all}}$ (red) and $V_{\text{sel}}$ (blue) for $\langle N_\text{PU} \rangle = 40$ .}
\end{figure}

Plots in Fig.~\ref{fig:roc1} summarize our result for $\langle N_\text{PU} \rangle = 40$.  In the top frame we show the background efficiency ($\epsilon_B$) as a function of the signal efficiency ($\epsilon_S$). The three curves in the plot refer to three different BDTs, constructed out of the variables in $V_\text{c}$ (solid black),  $V_\text{all}$ (solid red), and $V_\text{sel}$ (solid blue). In the same plot also show the results obtained using $\langle N_\text{PU} \rangle = 20$  with dashed lines. The pile-up robustness is
conspicuous. We also notice that one can drastically reduce the background (by factors of order $10$) while keeping the same signal acceptance when us use $V_\text{all}$. The effectiveness of the $\xi$ variables is more evident for the solid blue plot: where we show that keeping \emph{only} one conventional variable (namely $M_\text{eff}$), one can easily outperform the BDT with all conventional variables. 
In fact, we find that in our MVA $\xi(5)$ plays the most dominant role in the separation, followed by  $\xi(1)$, and then other variables (in the order, $p_T^J$, $H_{T}$, $M_\text{eff}$, $\xi(0.5)$, $\met$). This is encouraging -- $\xi$ variables not only improve $S/B$, these are actually more powerful than any others on our list of conventional variables. 

One  concern, however, remains --  whether or not the large purity of the sample comes at the cost of reducing statistical significance. We address this concern in the bottom frame of Fig.~\ref{fig:roc1}, where we plot relative improvements in the discovery potentials. To be specific, we define the improvement factor $\mathcal{I}$ to be   
\begin{equation}
\mathcal{I}_a \equiv \frac{\left( \epsilon_S/\sqrt{\epsilon_B} \right)_{a}}{\left( \epsilon_S/\sqrt{\epsilon_B} \right)_{V_\text{c}} }\, , \qquad a \ \in \ \{ V_\text{all}, V_\text{sel}  \}\, ,
\label{eq:imp_defn}
\end{equation}
where, the suffixes refer to the set of variables we use  to calculate the signal and background efficiencies. Again, the solid red and blue lines show the improvement factor if the BDT uses variables in $V_\text{all}$ and $V_\text{sel}$ respectively, vs. the BDT using only the conventional variable for $\langle N_\text{PU} \rangle = 40$.

The plots in Fig.~\ref{fig:roc1} sufficiently demonstrate the efficacy of $\xi$ as a discriminating variable.
\begin{table}[h]
\setlength{\tabcolsep}{6pt}
\renewcommand{\arraystretch}{1.2}
\begin{center}
\begin{tabular}{| l |c|c| c|}
\hline
Sample Cutflow & $\epsilon_S$ &  $\epsilon_B$ &  $\frac{S/B}{\left(S/B\right)_\text{C}}$ \\   
\hline
\hline
BDT with all variables in $V_C$ &
	0.3 & 0.010 & 1 \\
\hline
$\xi(1)>18$, $\xi(5)>3$, &  
	\multirow{2}{*}{0.3} & \multirow{2}{*}{0.007} & \multirow{2}{*}{1.43} \\
$\met>135\gev$, $M_\text{eff} > 200\gev$ & &  & \\
\hline
\end{tabular}
\caption{\label{table:effs} Comparison of performances of a simple cut-flow (using $\xi$) with optimised MVA using only conventional variables for a fixed $\epsilon_B = 30\%$.}
\end{center}
\end{table}
However, in order to bring home the point we additionally show a sample cut-flow involving $\xi$ in Table.~\ref{table:effs}. We find that a simple cut on $\xi(1)$, $\xi(5)$, $\met$, and  $M_\text{eff}$ can easily improve $S/B$ by  more than $40\%$, for fixed $\epsilon_S=30\%$. We emphasize that this demonstrated performance enhancement is not restricted to this specific $\epsilon_S$. One can easily find sample cut-flows that outperform the optimized MVA using all the conventional variables, for the full range of $\epsilon_S$.

It is tempting to check whether $\xi$ can still remain useful in a far more challenging spectrum than what we have used. For this we introduce a ``super-compressed" scenario, where we use the same topology as before but with a degree of compression $\Delta = 1/75$. In particular, we use $m_A-m_B = m_B-m_C= m_C-m_D = m_D-m_E= 5\gev$, with $m_A = 1.5\tev$. Note that we expect $\xi$ to be less effective. This high degree of compression results in much softer partons from the hard interaction which, in turn, give rise to even softer hadrons. A significant number of tracks would not even be registered at the tracker level.  We notice that the introduction of $\xi$ still manages to improve the signal significance by $20$-$30\%$ for the ``super-compressed'' case. Not surprisingly, we find that $\xi(1)$ manages to separate signal events more than $\xi(5)$. Additionally, this observation lets us speculate that if, indeed,  NP is found in the monojet+$\met$ channel, $\xi(a)$ can play a big role in disentangling the details of decay topology. 

Before concluding, let us note that one concern remains about whether the distributions of $\xi$ for the background events can be estimated reliably. In this article we rely on Monte Carlo, and even though we agree that it may  not give very good measurements of $\xi$, there are much better ways available to estimate these. For example, one can identify $Zj$ events where $Z$ decays to muons, which subsequently get reconstructed to yield $Z$-mass. The distributions of soft tracks seen from these events (after the removal of tracks associated with the jet and the muons) give a reliable measure of $\xi$ in  the monojet + $\met$ channel for background due to $Zj$.    


\section*{Acknowledgements}
Part of this was completed in the Gaggle cluster at TIFR. TSR was supported in part by the SERB Early Career Research Award. The authors thank Adam Martin for his careful reading of the draft.


\bibliography{References}

\end{document}